\documentclass[%
pre,
 amsmath,amssymb,
 reprint,%
]{revtex4-1}

\usepackage{graphicx}
\usepackage{dcolumn}
\usepackage{bm}

\usepackage[utf8]{inputenc}
\usepackage[T1]{fontenc}
\usepackage{mathptmx}
\usepackage{etoolbox}

\makeatletter
\def\@email#1#2{%
 \endgroup
 \patchcmd{\titleblock@produce}
  {\frontmatter@RRAPformat}
  {\frontmatter@RRAPformat{\produce@RRAP{*#1\href{mailto:#2}{#2}}}\frontmatter@RRAPformat}
  {}{}
}%
\makeatother
\begin{document}

\preprint{AIP/123-QED}

\title[Long-Pulse Fast Ignition in MagLIF]{Long-Pulse Fast Ignition in MagLIF}

\author{Benjamin Wang}

\author{Henry Fetsch}

\author{Nathaniel J. Fisch}
 
\affiliation{Department of Astrophysical Sciences, Princeton University, Princeton, NJ
}

\email{bwwang@pppl.gov}

\date{\today}

\begin{abstract}

The fast ignition paradigm for inertial confinement fusion (ICF) allows for extremely high gains but requires fuel to be heated very quickly to outpace hotspot disassembly and energy losses. This demands lasers with high power and intensity, posing engineering challenges that have called into question the fundamental practicality of fast ignition. Magnetized liner inertial fusion (MagLIF) circumvents these problems through its large-aspect-ratio cylindrical geometry and strong axial magnetic fields that allow for ignition at lower areal densities. Furthermore, MagLIF’s large aspect ratio and higher yields relax other constraints on energy deposition and repetition rate while its axial magnetic fields can be used to collimate ignitor electrons and thereby increase allowed standoff distance and save on ignitor energy. This tremendous overall relaxation of the engineering constraints that have historically limited the practicality of fast ignition suggests that the paradigm may be considerably more viable in a MagLIF context.
\end{abstract}

\maketitle

\section{\label{sec:intro}Introduction}

Fast ignition is a paradigm for inertial confinement fusion (ICF) which separates the compression and heating stages of ICF by using an ignitor that rapidly heats a section of pre-compressed fuel to spark the fusion reaction. This allows for compression at low entropy and consequently extremely high gains, even to the point where advanced fuels become viable \cite{Tabak01042006}. The challenge is that fast ignition requires a very high ignitor power when implemented in laser ICF, with a standard density of $300\text{ g/cm}^3$ requiring an energy of around $\gtrsim 20\text{ kJ}$ to be delivered within about $\lesssim 20\text{ ps}$ \cite{Tabak01042006}, corresponding to a petawatt-scale power $\gtrsim 10^{15}\text{ W}$. If the ignitor pulse could be lengthened while maintaining low total energy delivered, it would greatly relax the engineering constraints on the implementation of fast ignition.

The magnetized liner inertial fusion (MagLIF) scheme for ICF offers a pathway to this relaxation. MagLIF implodes a cylindrical metal liner by running a large axial current through it, simultaneously amplifying an applied seed field by flux compression, a technique known to compress seed magnetic fields of <100 T up to kiloteslas in experiments \cite{PhysRevLett.103.215004,PhysRevLett.107.035006} and tens of kiloteslas in simulations \cite{10.1063/1.3333505,10.1063/1.4816813,PhysRevLett.108.025003}. These magnetic fields suppress thermal conduction losses \cite{PhysRevLett.107.035006} and magnetize fusion alphas \cite{PhysRevLett.113.155004}, enabling ignition at much lower areal densities \cite{10.1063/1.3333505,M.M.Basko_2000}. High-current drivers, such as the proposed Z800 which could deliver $65\text{ MA}$ in $113\text{ ns}$ \cite{PhysRevSTAB.18.110401,10.1063/1.4941100}, can be used in conjunction with ice-burning MagLIF targets to achieve simulated yields \ on the order of $\sim 7\text{ GJ}$ \cite{PhysRevLett.108.025003,10.1063/1.4941100} and simulated central densities and cold fuel densities of up to $10\text{ g/cm}^3$ and $250\text{ g/cm}^3$ respectively \cite{PhysRevLett.108.025003}.

Fast ignition has not been previously considered in MagLIF because the low hotspot densities achieved on the Z machine ($\lesssim 1\text{ g/cm}^3$) are unsuitable for ignitor energy deposition and because MagLIF can already ignite at much lower pressures than laser ICF, seemingly obviating the need for fast ignition. However, as we will see, fast ignition can still allow for significantly reduced stagnation pressures compared to compression heating MagLIF configurations, with high yield MagLIF becoming viable even at stagnation pressures on the order of those achievable by the Z machine.

The properties of MagLIF implosions offer a solution to the central challenge of fast ignition, namely that a hot spot generated suddenly in an isochoric background will expand quickly, losing energy and eventually becoming too diffuse to ignite. 
The cylindrical target configuration is conducive to cylindrical hot spots, whose volumes scale with radius $R$ as $R^2$ rather than the spherical $R^3$ of laser ICF, leading to a slower decrease in fuel density. Moreover, the ability of magnetized fuel to ignite at lower areal densities mitigates the problem of fuel rarefaction due to expansion. 
MagLIF additionally offers better energy confinement per ion in fast-ignition hot spots: thermal conduction, which is the dominant energy loss mechanism for long-pulse fast ignition in laser ICF, is effectively eliminated by the magnetic fields in MagLIF while bremsstrahlung and PdV losses are kept manageable by the lower densities and cylindrical geometry respectively. The cumulative effect is that long-pulse fast ignition becomes viable in MagLIF, with a 100 ps, 5.19 kJ ignitor pulse achieving ignition in the model described below.

This work is organized as follows. First, in section \ref{sec:expansion_model}, we obtain an expression for the expansion speed of the hotspot, which is important because it governs PdV losses, hotspot disassembly, and evolution of the density and magnetic field within the hotspot. Next, in section \ref{sec:energy_model}, we come up with a simple model to describe the energy dynamics in a MagLIF fast ignition hotspot. Finally, in sections \ref{sec:results} and \ref{sec:discussion} we use this model to estimate how much ignitor energy it would take to achieve ignition and discuss the implications of these results.

\section{\label{sec:expansion_model}Hotspot Expansion Model}

\begin{figure}[h]
\includegraphics[width=8cm]{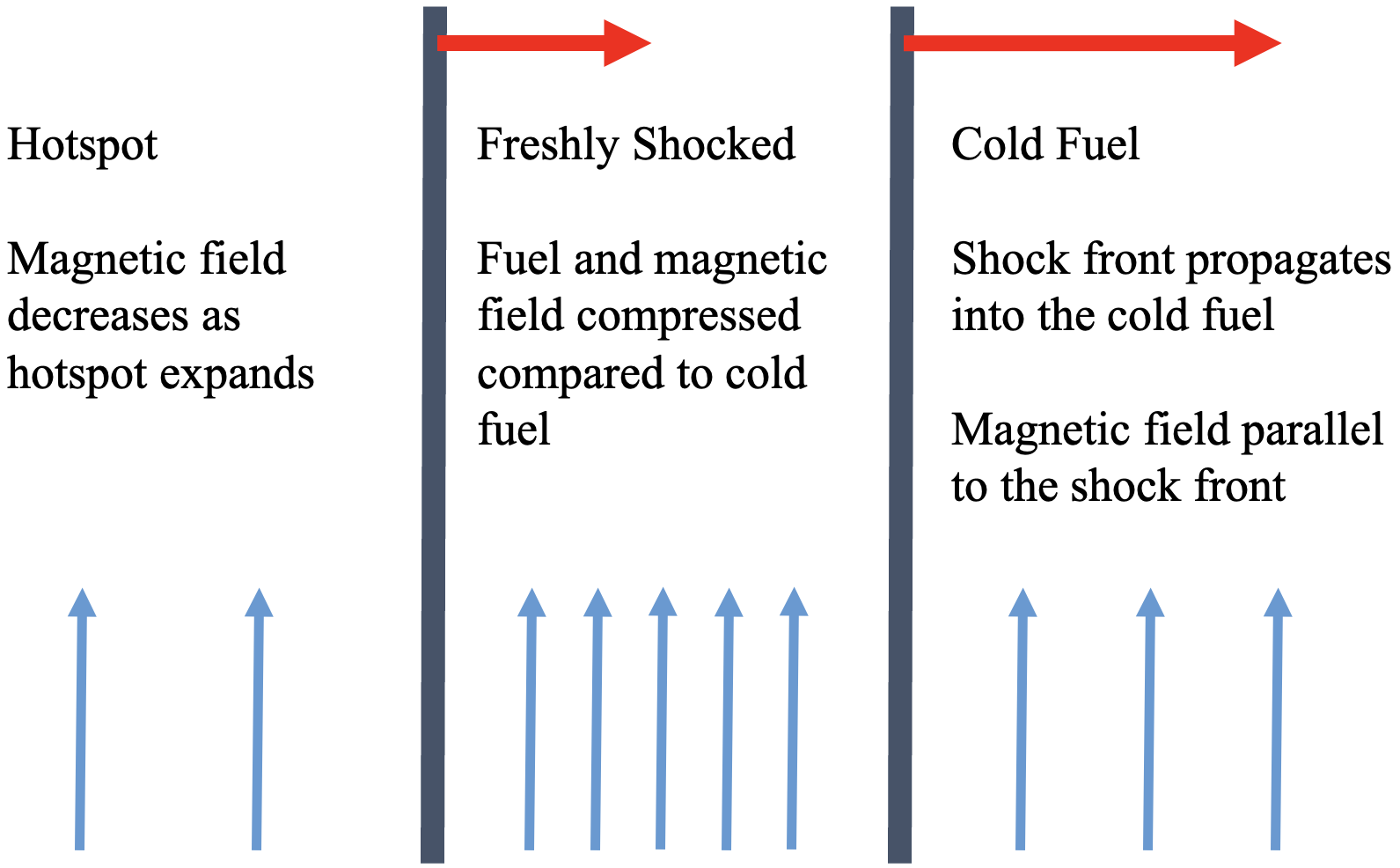}
\caption{The three regions in our model.}
\label{fig:0d_model}
\end{figure}

Consider a long cylindrical hot spot expanding radially. The expansion speed can be described by the following model, which may be considered to be a generalization of the classic model where an unmagnetized hot spot expands at the speed of material behind a strong shock \cite{Tabak01042006}. Let there be three regions, a hotspot (subscript 1), a freshly shocked region (subscript 2), and cold fuel (subscript 0) of density $\rho_0$, see Fig.\ \ref{fig:0d_model}.

Define the boundary of the hotspot so that there is no mass flux between the hotspot and the freshly shocked region. There is a shock discontinuity propagating into the cold fuel between the freshly shocked region and the cold fuel. Let there be a magnetic field parallel to the shock front which has strength $B_0$ in the cold fuel. Assume a cylindrical geometry and that the hotspot has expanded by a factor of $x$ so that the magnetic field inside the hotspot has strength $B_0/x^2$. Let the compression ratio between the freshly shocked region and the cold fuel be $r$ so that $\rho_2=r\rho_0$ which from boundary conditions implies $B_2=rB_0$. Define cold fuel pressures $P_0=aP_1$ and $\frac{B_0^2}{2\mu_0}=bP_1$. Now boundary conditions (effectively pressure balance) give us
\begin{equation}
    P_1+\frac{B_0^2}{2\mu_0 x^4}=P_2+\frac{r^2 B_0^2}{2\mu_0}\Rightarrow P_2=\left(1-b\left(r^2-\frac{1}{x^4}\right)\right)P_1.
\end{equation}

Shifting into the shock frame and neglecting the electric field due to non-relativistic speeds, let $U_0$ be the speed of the cold fuel approaching the shock and let $U_2$ be the speed of the freshly shocked fuel leaving the shock. Assuming an ideal gas $\gamma=\frac{5}{3}$, we apply conservation of mass, momentum, and energy, respectively, to get
\begin{equation}
    \rho_2 U_2=\rho_0 U_0\Rightarrow rU_2=U_0,
\end{equation}
\begin{equation} \label{consp}
    r\rho_0 U_2^2+P_2+\frac{r^2 B_0^2}{2\mu_0}=r^2 \rho_0 U_2^2+aP_1+\frac{B_0^2}{2\mu_0},
\end{equation}
\begin{equation} \label{consE}
    \frac{1}{2}r\rho_0 U_2^2+\frac{5}{2}P_2+\frac{r^2 B_0^2}{\mu_0}=r\left(\frac{1}{2}r^2 \rho_0 U_2^2+\frac{5}{2}aP_1+\frac{B_0^2}{\mu_0}\right).
\end{equation}
Rewriting all pressures in Eq.\ \ref{consE} in terms of $P_1$,
\begin{equation}
    \left(\frac{5}{r}\left(1-b\left(r^2-x^{-4}\right)\right)+4(r-1)b-5a\right)P_1=(r^2-1) \rho_0 U_2^2,
\end{equation}
which yields
\begin{equation}
    P_1=\frac{r(r^2-1)\rho_0 U_2^2}{5(1-b(r^2-x^{-4}))+4r(r-1)b-5ra}.
\end{equation}
Using this to rewrite all pressures in Eq.\ \ref{consp} in terms of $\rho_0 U_2^2$ and then dividing out by $\rho_0 U_2^2$, we get an equation for the compression ratio $r$,
\begin{equation}
    r+\frac{r(r^2-1)(1-b(1-x^{-4})-a)}{5(1-b(r^2-x^{-4}))+4r(r-1)b-5ra}=r^2.
\end{equation}
In the cold, unmagnetized limit $a=b=0$, we recover the ideal gas compression limit $r=4$.

\begin{figure}[h]
\includegraphics[width=8cm]{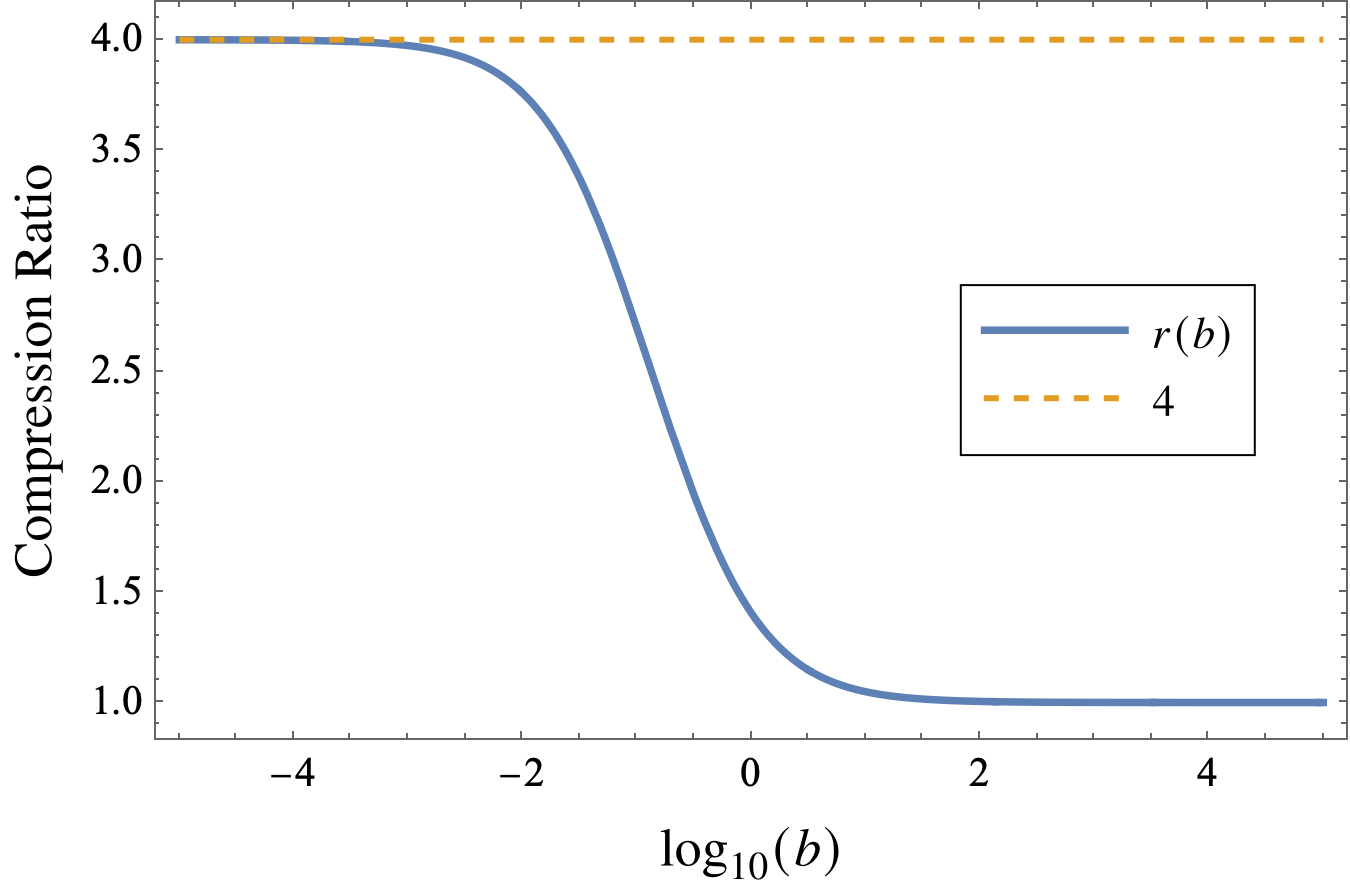}
\caption{Compression ratio vs.\ magnetic pressure for radial expansion ratio $x=1$ and cold fuel pressure $a=0$.}
\end{figure}

Rewrite Eq.\ \ref{consp} as
\begin{equation}
    r(r-1)\rho_0 U_2^2=(1-b(1-x^{-4})-a)P_1
\end{equation}
which yields a shock speed of
\begin{equation}
    U_0=rU_2=\sqrt{\frac{r(1-b(1-x^{-4})-a)P_1}{(r-1)\rho_0}}
\end{equation}
and a hotspot expansion speed of
\begin{equation}
\label{hotspot_speed}
    U_0-U_2=(r-1)U_2=\sqrt{\frac{(r-1)(1-b(1-x^{-4})-a)P_1}{r\rho_0}}.
\end{equation}

\begin{figure}[h]
\includegraphics[width=8cm]{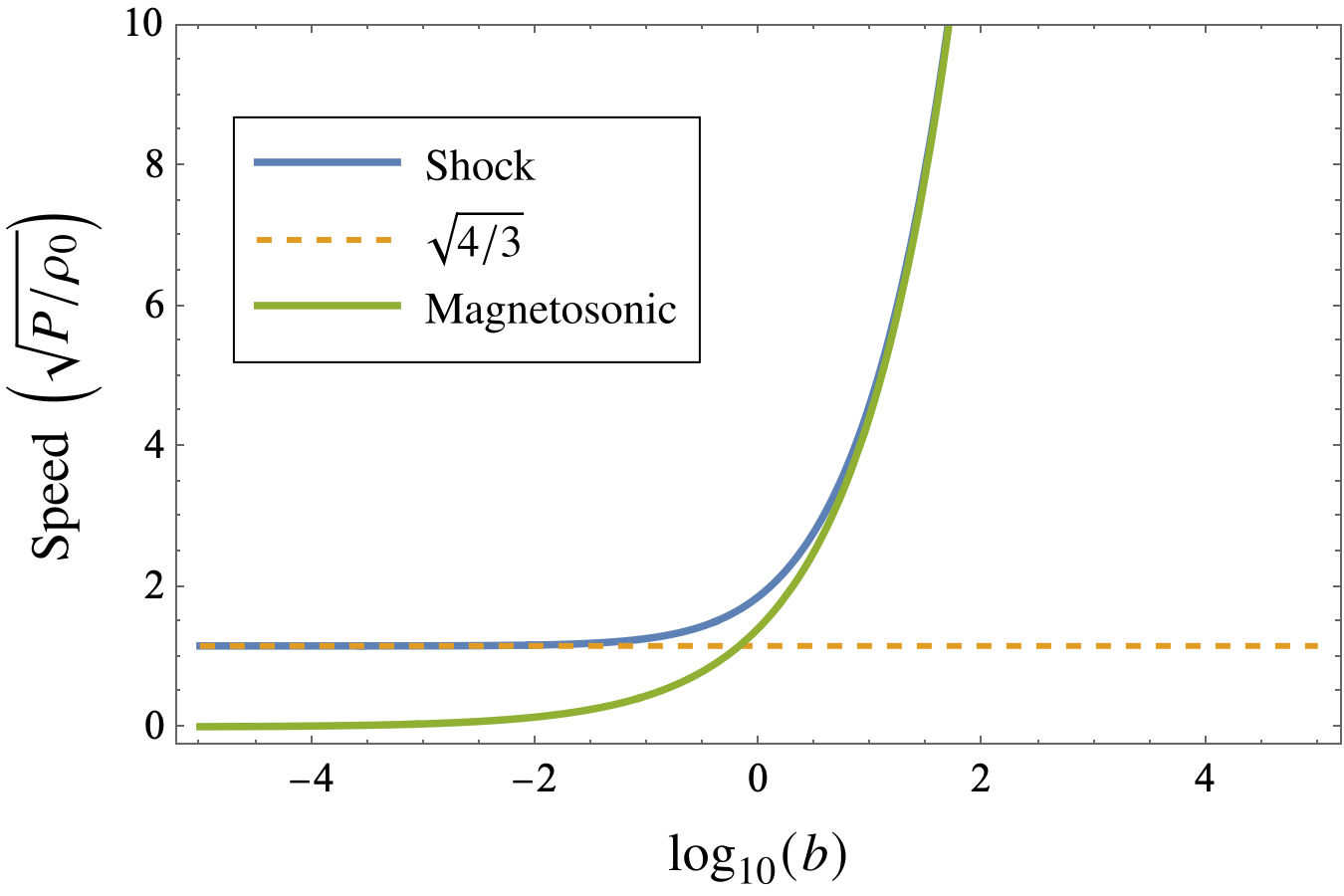}
\caption{Shock speed vs.\ magnetic pressure for radial expansion ratio $x=1$ and cold fuel pressure $a=0$. In order for the front to be a shock, it must travel faster than the sound (magnetosonic) speed in the cold fuel.}
\end{figure}

\begin{figure}[h]
\includegraphics[width=8cm]{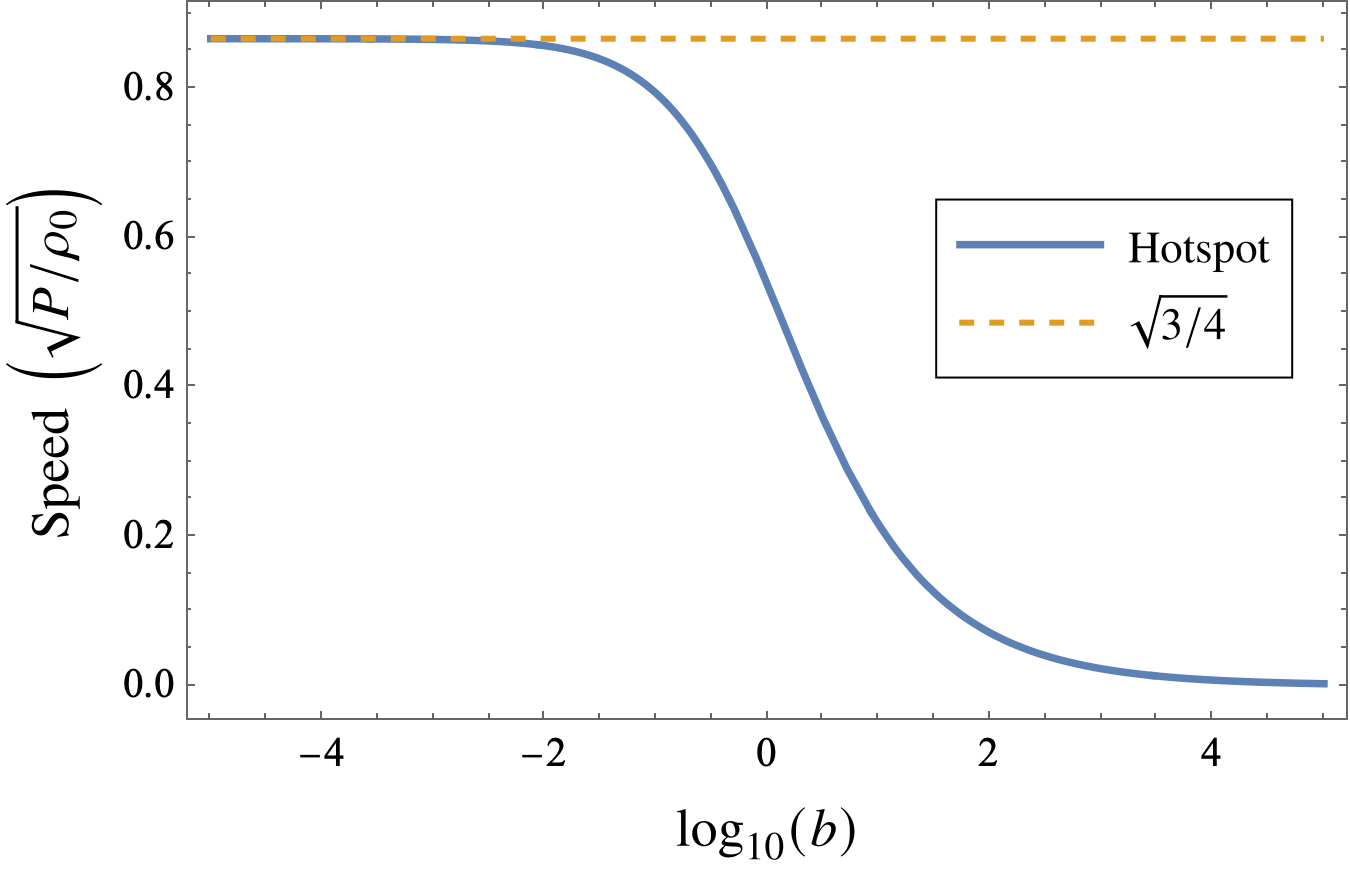}
\caption{Hotspot speed vs.\ magnetic pressure for radial expansion ratio $x=1$ and cold fuel pressure $a=0$.}
\end{figure}

\begin{figure}[h]
\includegraphics[width=8cm]{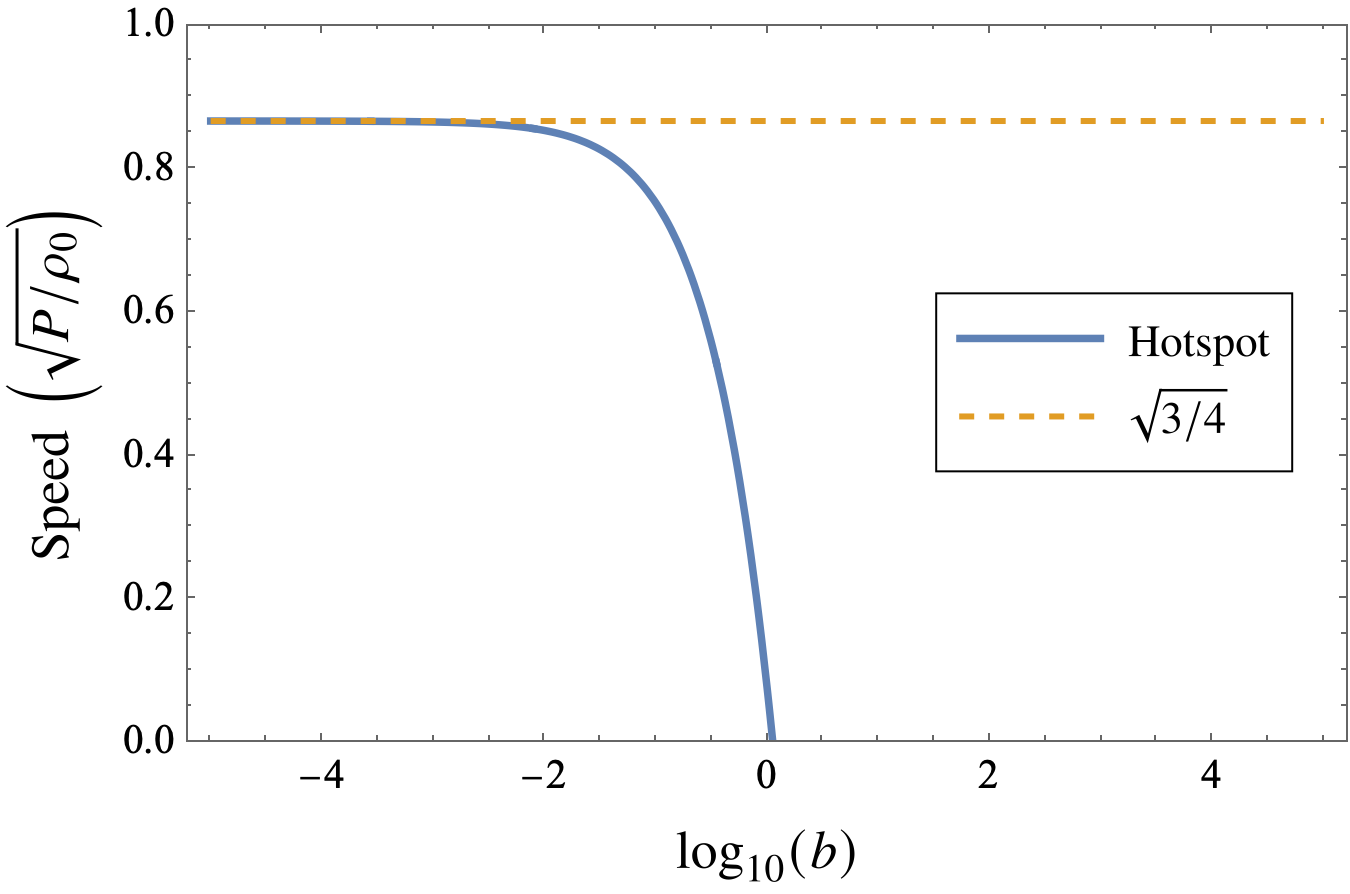}
\caption{Hotspot speed vs.\ magnetic pressure for radial expansion ratio $x=1.75$ and cold fuel pressure $a=0$. Hotspot speed goes to zero when total pressure (thermal plus magnetic) in the hotspot equals total pressure in the cold fuel.}
\label{fig:hotspot_speed2}
\end{figure}

We see that the magnetic field slows down the hotspot expansion through magnetic pressure and speeds up the shock propagation by stiffening the cold fuel. Note that the increase in the shock speed does not affect anything because what it effectively does is cause the shock to hit the infalling liner sooner, which is an effect that we are not considering. 

\section{\label{sec:energy_model}Energy Dynamics Model}

\begin{figure}[h]
\includegraphics[width=8cm]{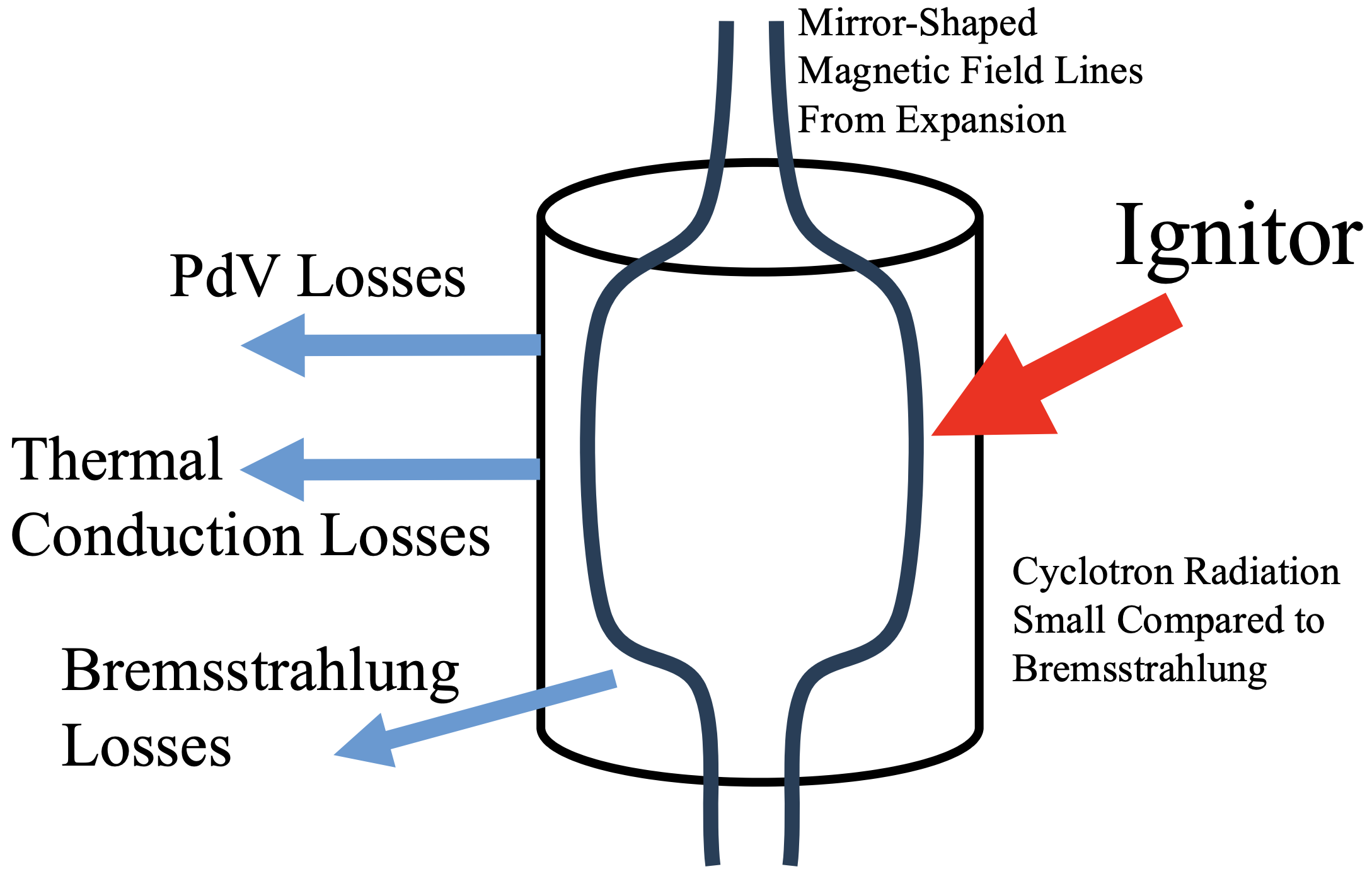}
\caption{Energy dynamics of the hotspot.}
\label{fig:energy_dynamics}
\end{figure}

We would like to estimate how the required energy deposition to achieve ignition scales with input parameters such as magnetic field strength, initial hotspot radius, and initial fuel density. Consider a simplified model where we start with a large aspect ratio cylindrical hotspot (so that only radial expansion and radial thermal conduction losses are significant) initially at equal density with the surrounding cold fuel in a uniform axial magnetic field. Assume that the hotspot expands radially with ideal gas $\gamma=5/3$ at the speed calculated in Eq.\ \ref{hotspot_speed}. The hotspot experiences energy input from the ignitor, which we assume to go purely into the electrons, and energy losses from PdV work, thermal conduction, and bremsstrahlung, see Fig.\ \ref{fig:energy_dynamics}. We assume a 50-50 DT fuel mix and neglect the cold fuel pressure throughout this work.

First, we use Braginskii's expression for the classical heat flux in subsection \ref{classical} to estimate a reasonable radial temperature profile inside the hotspot in subsection \ref{htp}. Next, we add nonlocal corrections to the thermal conduction in subsection \ref{nonlocal} and add PdV and bremsstrahlung losses in subsection \ref{pdvbrem}. Finally, we add electron-ion temperature equilibration in subsection \ref{tempevo} and explain what ignitor pulse shape we used for our "optimal" ignitor runs in subsection \ref{pulseshape}.

\subsection{\label{classical}Classical Thermal Conduction}

We use cgs units and keV for temperature. The classical heat flux per unit area \cite{Atzeni_Meyer-Ter-Vehn_2004} is $q_{\parallel C}=-\kappa_{e\parallel} \nabla T_e$ where $\kappa_{e\parallel}=A_e T_e^{5/2}/\ln{\Lambda}$ with $A_e=9.5\times 10^{19}\text{ erg s}^{-1}\text{ cm}^{-1}\text{ keV}^{-7/2}$. Radial thermal conduction is perpendicular to the magnetic field, so we multiply by the classical heat transport ratio \cite{1965RvPP....1..205B} for $Z=1$,
\begin{equation}
    \kappa_{e \perp}=\frac{4.664\chi^2+11.92}{3.1616(\chi^4+14.79\chi^2+3.7703)}\kappa_{e \parallel},
\end{equation}
where $\chi=\omega_e\tau_e$ is the Hall parameter.

A heat flux of $q_e$ results in Nernst advection of the magnetic field with a speed of $v_N\sim 2q_e/3n_e kT_e$ \cite{PhysRevLett.53.262}. Therefore, for a hotspot of radius $R_h$ we estimate the Nernst velocity as $v_N\sim \frac{2\kappa_{e\perp}}{3R_h n_e k}$ and the Nernst advection timescale as $\tau_N\sim \frac{R_h}{v_N}\sim \frac{3R_h^2 n_e k}{2\kappa_{e\perp}}$. Plugging in $\rho_0=100\text{ g/cm}^3$, $B_0=30\text{ kT}$, $R_h=10$ \textmu m, and $T_e=10\text{ keV}$, we have $\tau_N\gtrsim 2000\text{ ps}$, which is much longer than the ignitor pulse length scale, so we neglect Nernst advection. Note that in this regime, the Nernst velocity actually decreases with temperature, because increasing temperature increases $\chi$ which decreases $\kappa_{e\perp}$.

\subsection{\label{htp}Hotspot Temperature Profile}

We estimate the radial temperature profile inside the hotspot as that which results in spatially uniform thermal conduction losses throughout the hotspot. This is reasonable because, assuming constant pressure throughout the hotspot, the density varies inversely with temperature, so spatially uniform energy losses in the hotspot preserve the temperature profile. For a cylindrical geometry, the volume enclosed by a given radius $R$ is proportional to $R^2$, so we have that the total heat flux through each radius $R$ is proportional to $R^2$, giving in the unmagnetized case
\begin{equation}
    R T_e^{5/2}\frac{dT_e}{dR}\sim R^2
\end{equation}
which for a central hotspot electron temperature $T_{eh}$ that drops to zero at the hotspot boundary $R=R_h$ has the solution
\begin{equation}
    T_e(R)=T_{eh}\left(1-\left(\frac{R}{R_h}\right)^2\right)^{2/7}.
\end{equation}
This temperature profile gives a heat flux per unit area out of the hotspot surface of
\begin{equation}
    q_{\parallel C}=-\frac{A_e T_{eh}^{5/2}}{\ln{\Lambda}}\frac{4}{7}\frac{T_{eh}}{R_h}
\end{equation}
and has a density-weighted average temperature of
\begin{equation}
    \overline{T}_e=\frac{\int_0^{R_h} RdR}{\int_0^{R_h} \frac{R}{T_e(R)}dR}\approx 0.714 T_{eh}.
\end{equation}

\begin{figure}[h]
\includegraphics[width=8cm]{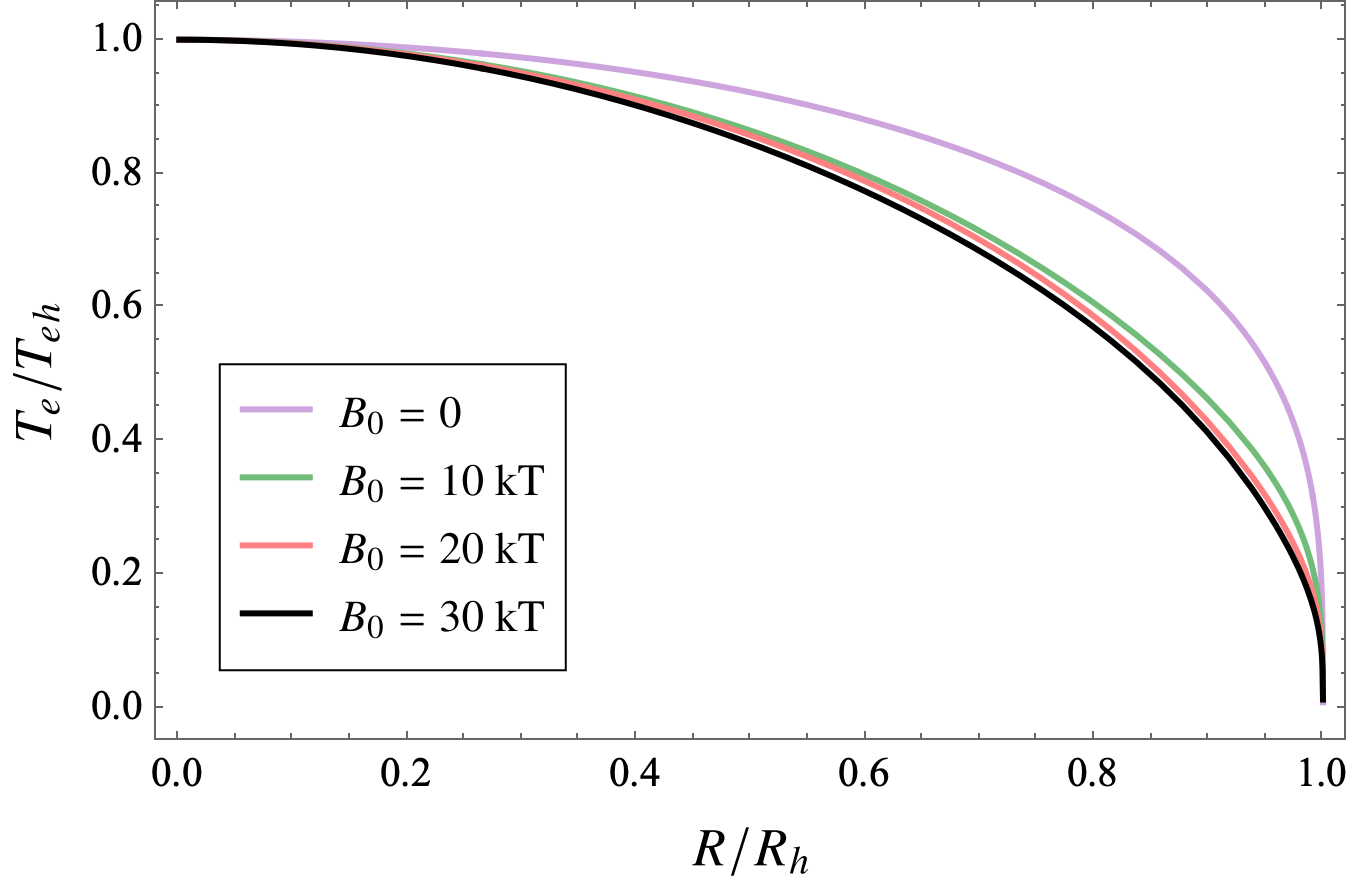}
\caption{Temperature vs.\ radius in the hotspot for various $B_0$ at $\rho_0=\text{50 g/cm}^3$, $\ln{\Lambda}=4$, and $T_{eh}=3\text{ keV}$.}
\label{fig:temperature_profile}
\end{figure}

In the magnetized case, because the magnetic field strength $B(R)$ at any radius $R$ is proportional to the density $n_e(R)$ by the frozen-in law, we have that the Hall parameter $\chi(R)$ and therefore the thermal conductivity $\kappa_{e\perp}(R)$ depend only on $T_e(R)$, $B_0$, and $n_0$, meaning that for each set $B_0$ and $n_0$ we can write $\kappa_{e\perp}$ as purely a function of temperature. Again assuming spatially uniform thermal conduction losses, we now numerically calculate the temperature profiles for each set $B_0$ and $n_0$ at a characteristic $T_{ehc}\equiv 3$ keV, see Fig.\ \ref{fig:temperature_profile}. We then use the shapes of these profiles to calculate the ratio $\overline{T}_e/T_{eh}$ and to evaluate the thermal conduction losses for an arbitrary $T_{eh}$ by approximating $q_{\perp C}(T_{eh})\approx q_{\perp C}(T_{ehc})\frac{\kappa_{e\perp}(\overline{T}_{eh})}{\kappa_{e\perp}(\overline{T}_{ehc})}\frac{T_{eh}}{T_{ehc}}$.

As can be seen in Fig.\ \ref{fig:temperature_profile}, adding a magnetic field results in a temperature profile which is more sharply peaked than in the unmagnetized case, lowering $\overline{T}_e/T_{eh}$ and giving an "advantage" to the magnetic field in the form of lower energy losses from the lower average temperature. This is not only reasonable but actually underestimates the true advantage of the magnetic field, because in reality the ignitor pulse intensity is peaked towards the center of the hotspot, and so with the magnetic field suppressing the redistribution of energy within the hotspot by thermal conduction, there will be a more sharply peaked temperature profile.

\subsection{\label{nonlocal}Nonlocal Thermal Conduction}

Due to the steep temperature gradients between the hotspot and the cold fuel, there is nonlocal heat transport where the electron mean free path becomes big relative to the temperature gradient length scale. To correct for this, we use a harmonic flux limiter (harmonic flux limiters fit Vlasov-Fokker-Planck simulation data much better than hard cutoffs \cite{10.1063/5.0225592}) with the expression
\begin{equation}
    q_{\parallel FL}=\frac{1}{\frac{1}{q_{\parallel C}}+\frac{1}{f_L q_F}}
\end{equation}
where $q_F=v_{\text{th}}nk_B T$ is the free-streaming limit and $f_L$ is the flux limiter, a constant multiple of the free-streaming limit that the heat flux caps out at. When the plasma is magnetized, the smallness of the electron gyroradius relocalizes the thermal conduction, so we have a correction factor \cite{10.1063/5.0225592} not only to the classical term but also to the flux limiter term, so that
\begin{equation}
    q_{\perp FL}=\min\left(\frac{1}{\frac{1}{q_{\perp C}}+\frac{\lambda_B}{\lambda_d}\frac{1}{f_L q_F}},q_{\parallel FL}\right)
\end{equation}
where $\lambda_d=\lambda_{ei}\sqrt{\frac{Z^2+0.24Z}{Z+4.2}}$, $\lambda_B=\frac{1}{\frac{1}{\lambda_d}+\frac{1}{r_L}}$, and $r_L=\frac{\lambda_{ei}}{\chi}$ is the electron Larmor radius. We use the values at the center of the hotspot to calculate $q_F$ and $\frac{\lambda_B}{\lambda_d}$. To estimate a flux limiter $f_L$ for each run we use the nonlocal Schurtz-Nicolaï-Busquet (SNB) model of thermal conduction \cite{10.1063/1.1289512,10.1063/5.0275673,10.1063/5.0005552,10.1063/1.5001079} as follows. For a run with number density $n_0$ and initial hotspot radius $R_0$, we consider a 1-D hyperbolic tangent temperature profile of $\frac{T_e(s)}{\text{keV}}=12.5+12.5\tanh\left(\frac{-s}{0.1 R_0}\right)$ and its corresponding density profile of $n(s)=0.714n_0\frac{25\text{ keV}}{T_e(s)}$. This temperature profile drops smoothly from 25 keV to zero in the span of about half a hotspot radius. We then calculate the SNB heat flux $Q(s)$ throughout space and take $f_L=\frac{\max(Q(s))}{q_F}$ with $q_F$ calculated at a temperature of 25 keV and a density of $0.714n_0$. 

\begin{figure}[h]
\includegraphics[width=8cm]{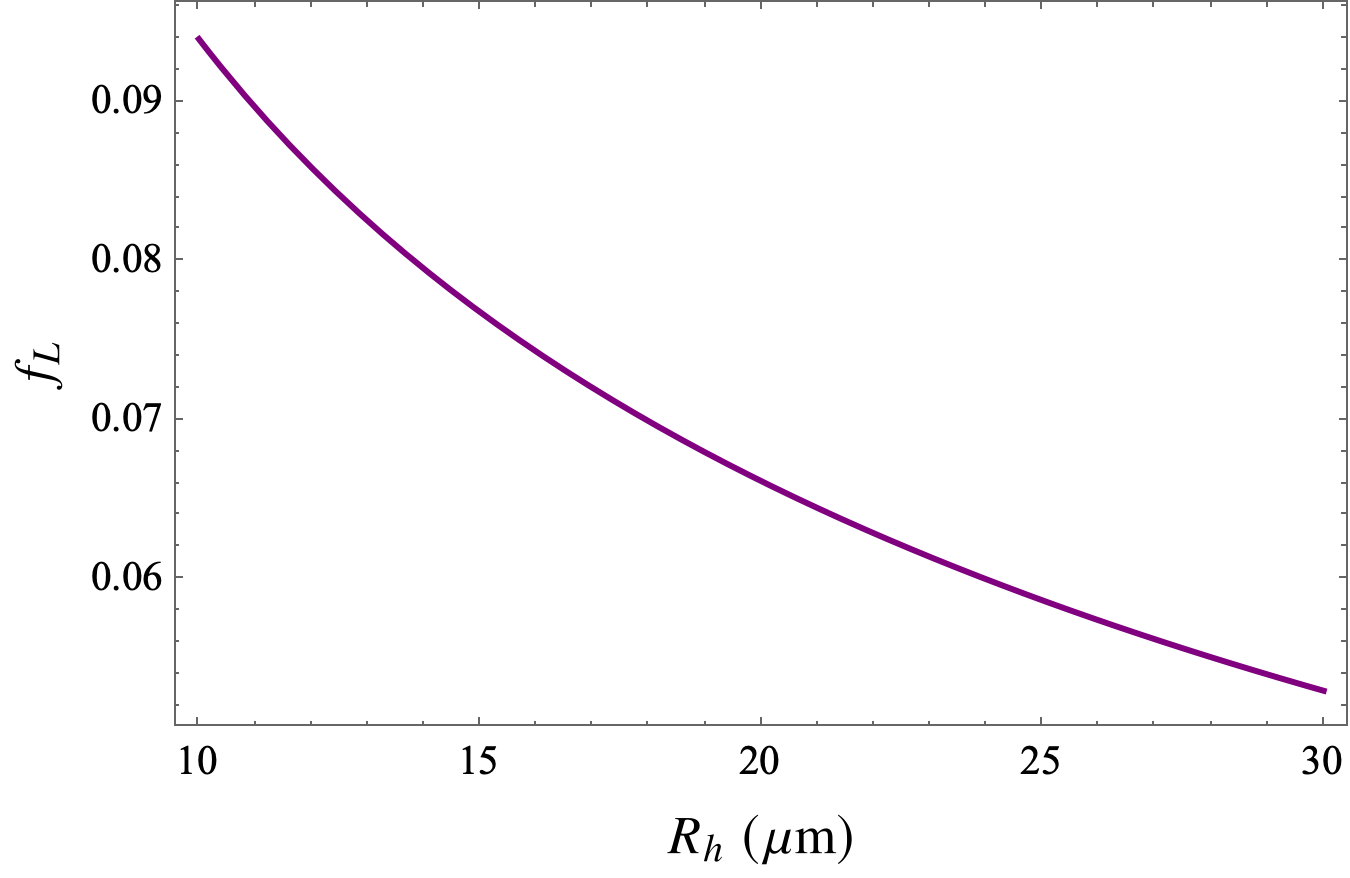}
\caption{Flux limiter vs.\ hotspot radius for $\rho_0=\text{300 g/cm}^3$ and $\ln{\Lambda}=3$.}
\label{fig:flux_limiter}
\end{figure}

Note that we have calculated a flux limiter using values representative of the starting configuration of the run; the effective flux limiter may actually increase throughout the run as the hotspot density decreases, this results in us having a small underestimation of thermal conduction losses.

\subsection{\label{pdvbrem}PdV and Bremsstrahlung Losses}

Assuming that ions and electrons have the same temperature profile shape, the PdV power per unit cylinder length for species $s$ is
\begin{equation}
    W_s=\frac{n_0}{x^2} k\overline{T}_s2\pi R_h\frac{dR_h}{dt}.
\end{equation}

We take the electron bremsstrahlung power radiated per unit volume \cite{Atzeni_Meyer-Ter-Vehn_2004} to be approximated by
\begin{equation}
    W_r=A_b\frac{\rho_0^2}{x^4}\overline{T}_e^{1/2}
\end{equation}
where $A_b=3.05\times 10^{23}\text{ erg cm}^3\text{ g}^{-2}\text{ s}^{-1}\text{ keV}^{-1/2}$ and assume that the hotspot is transparent to bremsstrahlung. We do not consider cyclotron radiation losses because even if we assume that the plasma is transparent to cyclotron radiation (in reality the cyclotron frequency is much less than the plasma frequency), the volumetric cyclotron radiation is still small compared to bremsstrahlung. An estimate of the volumetric cyclotron radiation power is given by ${P_c=6.21\times 10^{-13}B^2 n_e T_e \frac{\text{erg}}{\text{s T}^2\text{ eV}}}$ \cite{nrl_formulary}. Even in the worst case that we are considering, plugging in $\rho=50\text{ g/cm}^3$, $T_e=20\text{ keV}$, and $B=30\text{ kT}$ gives us a ratio of only $\frac{P_c}{W_r}=0.039$.

\subsection{\label{tempevo}Temperature Evolution}

We approximate the temperature equilibration rate \cite{nrl_formulary} between ions and electrons as
\begin{equation} \label{tempeq}
    \frac{d\overline{T}_i}{dt}=-\frac{d\overline{T}_e}{dt}=A_{ie}\frac{Z^2\frac{n_0}{x^2}\ln{\Lambda}}{m_i \overline{T}_e^{3/2}}\left(\overline{T}_e-\overline{T}_i\right)
\end{equation}
where $A_{ie}=1.7\times 10^{-37}\text{ g cm}^3\text{ keV}^{3/2}$. Each species $s$ has an ideal gas internal energy of $U_{int}=\frac{3}{2}N_0 k\overline{T}_s$
where $N_0$ is the number abundance $N_0\equiv\pi R_{0}^2 n_0$, so any power input or loss $W_s$ causes $T_{sh}$ to evolve as $\frac{dT_{sh}}{dt}=\frac{2}{3N_0 k}\frac{T_{sh}}{\overline{T}_s}W_s$.

\subsection{\label{pulseshape}Ignitor Pulse Shaping}

If we had an "optimal" ignitor such that we could perfectly control the electron temperature as a function of time, there would be two natural targets to optimize for: we could set the electron temperature to maximize the ion heating with respect to either the time or the hotspot expansion. If there were no hotspot expansion, then the optimal electron temperature would be held at $\overline{T}_e=3\overline{T}_i$ to maximize ${(\overline{T}_e-\overline{T}_i)/{\overline{T}_e^{3/2}}}$ in Eq.\ \ref{tempeq} and therefore heat the ions in as short a time as possible. However, because the hotspot does expand, we are better served by maximizing the ion heating with respect to the hotspot expansion, because causing the ion heating to outpace hotspot disassembly is essential and constitutes the more fundamental objective \cite{PhysRevE.108.045206}.

Because both the ion heating and the hotspot expansion rates depend on the electron temperature, it is nontrivial to solve this ignitor pulse shaping problem analytically even for the unmagnetized case as done in Fetsch and Fisch \cite{PhysRevE.108.045206} for a spherical geometry. The magnetized case is much more complicated, and so in our runs we obtain the optimal electron temperature by simply numerically maximizing the ion heating rate over the hotspot expansion rate at each timestep.

\section{\label{sec:results}Results}

\begin{figure}[h]
\includegraphics[width=8cm]{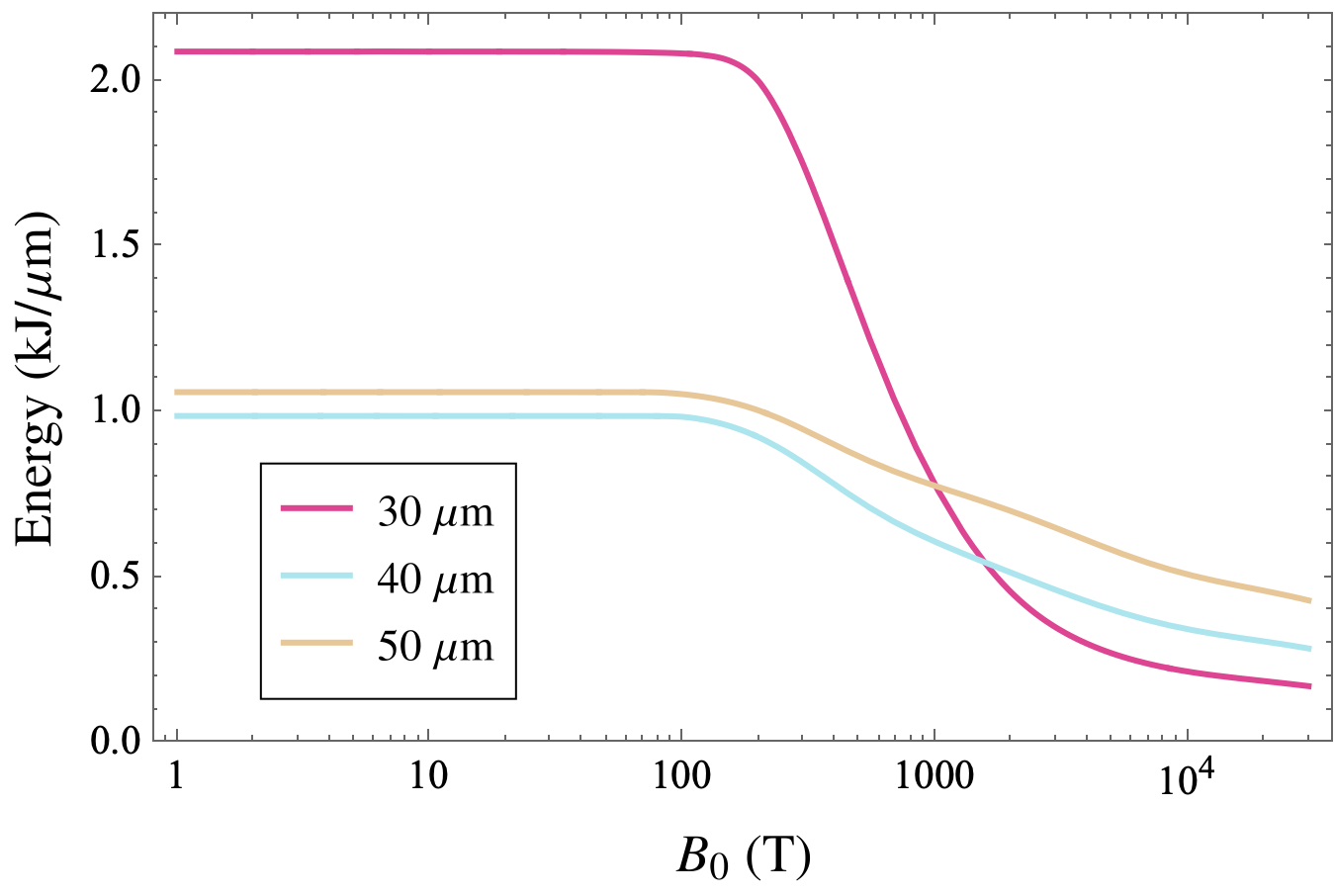}
\caption{Ignitor energy for optimal ignitor pulse vs.\ magnetic field strength for $\rho_0=\text{50 g/cm}^3$ and $\ln{\Lambda}=4$.}
\label{fig:energy_vs_B}
\end{figure}

Incorporating all of the energy dynamics shown in Fig.\ \ref{fig:energy_dynamics}, we track the ion and electron temperatures by integrating the energy inputs and outputs over time. First, we try to estimate how the energy required for ignition scales with magnetic field strength by considering an "optimal" ignitor pulse, meaning that at every timestep we set the electron temperature to maximize the ion heating per hotspot expansion. Using a density of $\rho_0=50\text{ g/cm}^3$, $\ln{\Lambda}=4$, and a negligible initial ion temperature of $\overline{T}_{i0}=100\text{ eV}$, we track the total energy input used to hit a central ion temperature of $10\text{ keV}$ versus magnetic field strength for initial hotspot radii of 30, 40, and 50 microns. These results are plotted in Fig.\ \ref{fig:energy_vs_B}.

\begin{figure}[h]
\includegraphics[width=8cm]{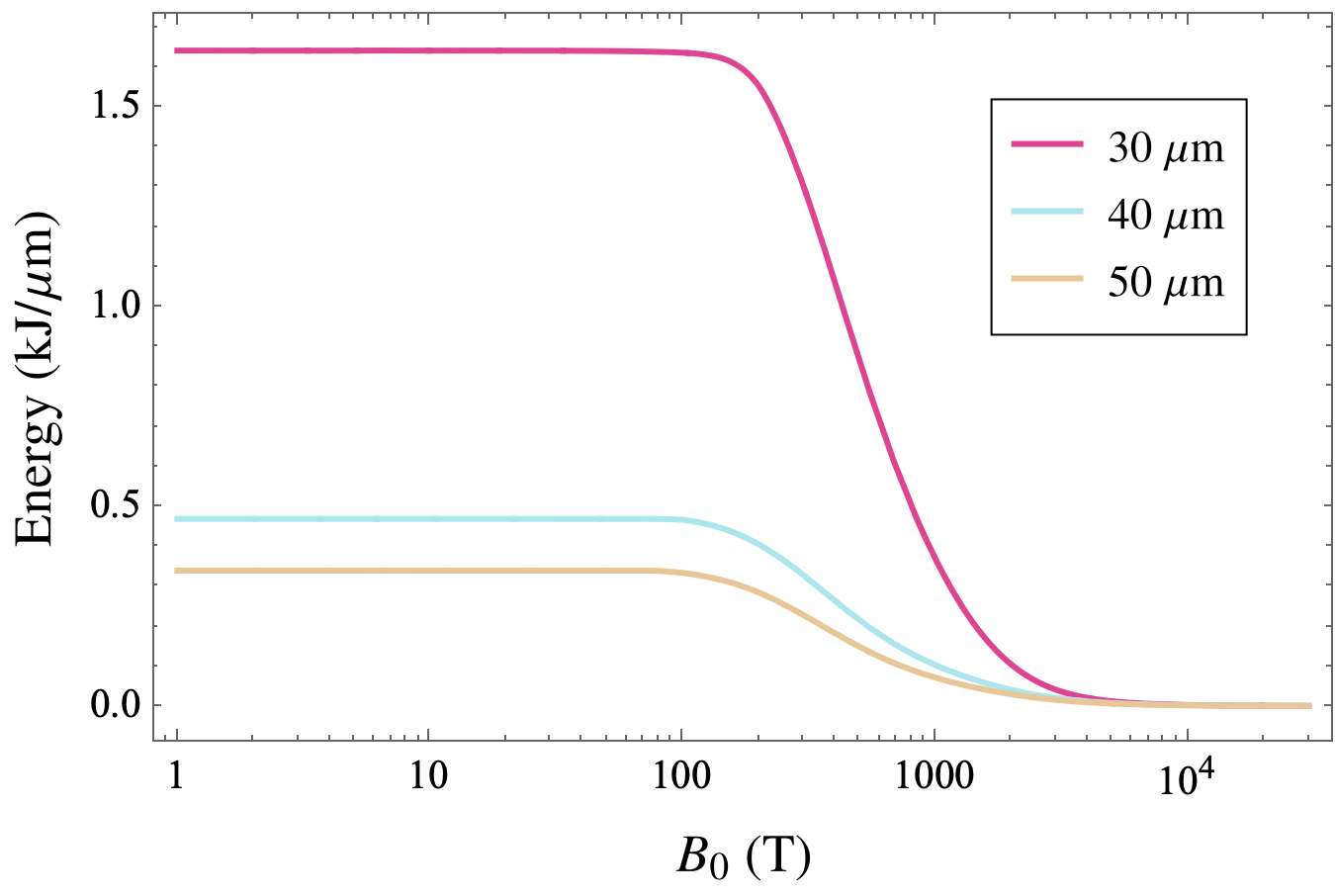}
\caption{Thermal conduction losses for Fig.\ \ref{fig:energy_vs_B} vs.\ magnetic field strength for $\rho_0=50\text{ g/cm}^3$ and $\ln{\Lambda}=4$.}
\label{fig:thermal_cond_losses}
\end{figure}

The energies in Fig.\ \ref{fig:energy_vs_B} are higher than is actually required for ignition because ignitor pulses which are optimized to strictly maximize the rate of ion heating are not the most energy efficient. However, these results still do provide some insight, because we see a much more dramatic decrease in the energy used for the 30 micron case than for the 40 or 50 micron cases as we increase the magnetic field. This is because the 30 micron case has a much longer pulse length since it starts at a smaller hotspot radius so expansion is more punishing, meaning that  the optimal pulse is less aggressive because it needs to worry more about limiting expansion rate. The energy lost to thermal conduction increases with pulse length and becomes the dominant energy expenditure for long pulses. What the results in Fig.\ \ref{fig:energy_vs_B} demonstrate is that strong magnetic fields, which effectively eliminate thermal conduction losses (see Fig.\ \ref{fig:thermal_cond_losses}), make long pulses viable where they otherwise would not be.

\begin{figure}[h]
\includegraphics[width=8cm]{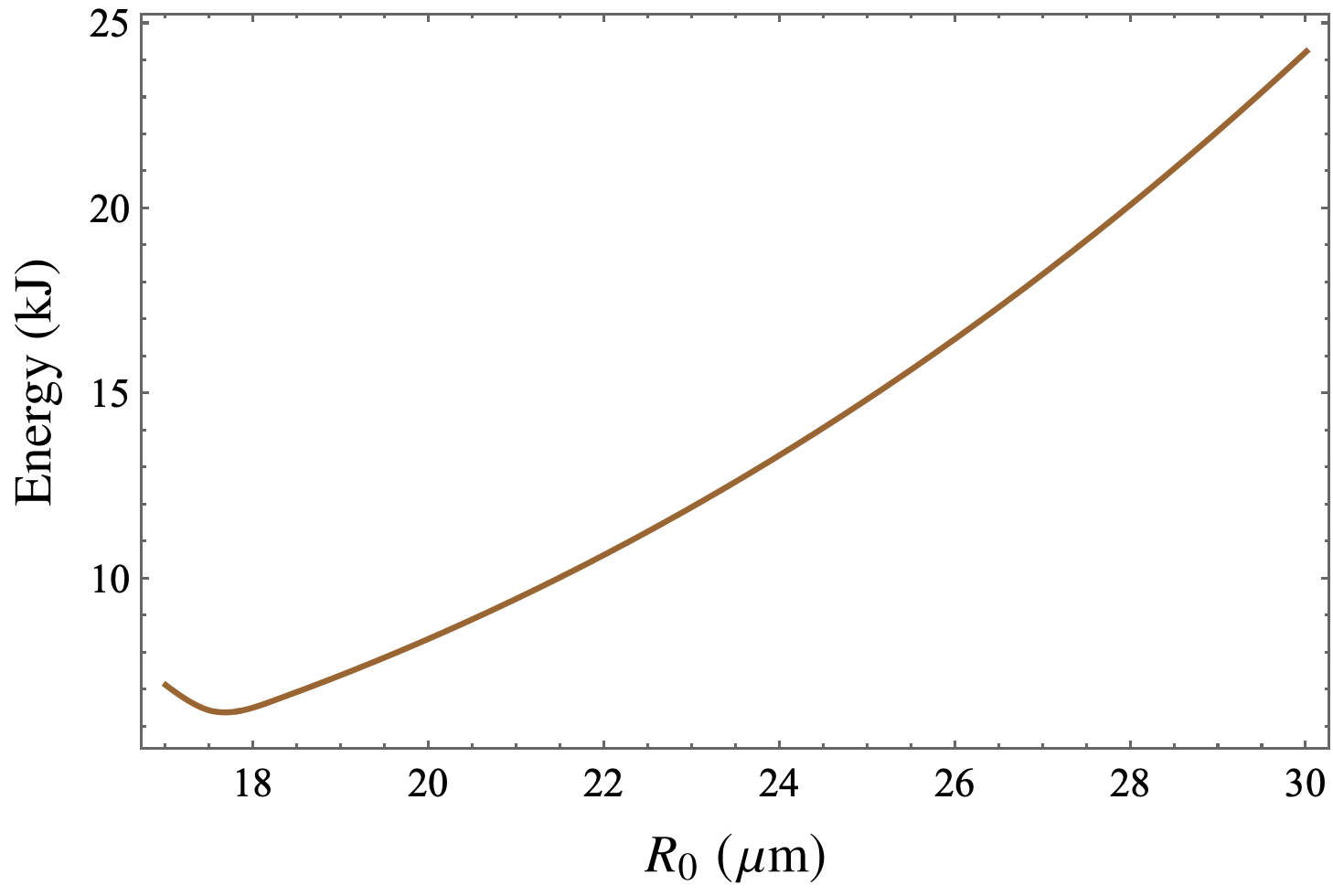}
\caption{Ignitor energy for 100 ps constant power pulse vs.\ initial hotspot radius for $\rho_0=\text{50 g/cm}^3$ and $\ln{\Lambda}=4$.}
\label{fig:50gcc}
\end{figure}

\begin{figure}[h]
\includegraphics[width=8cm]{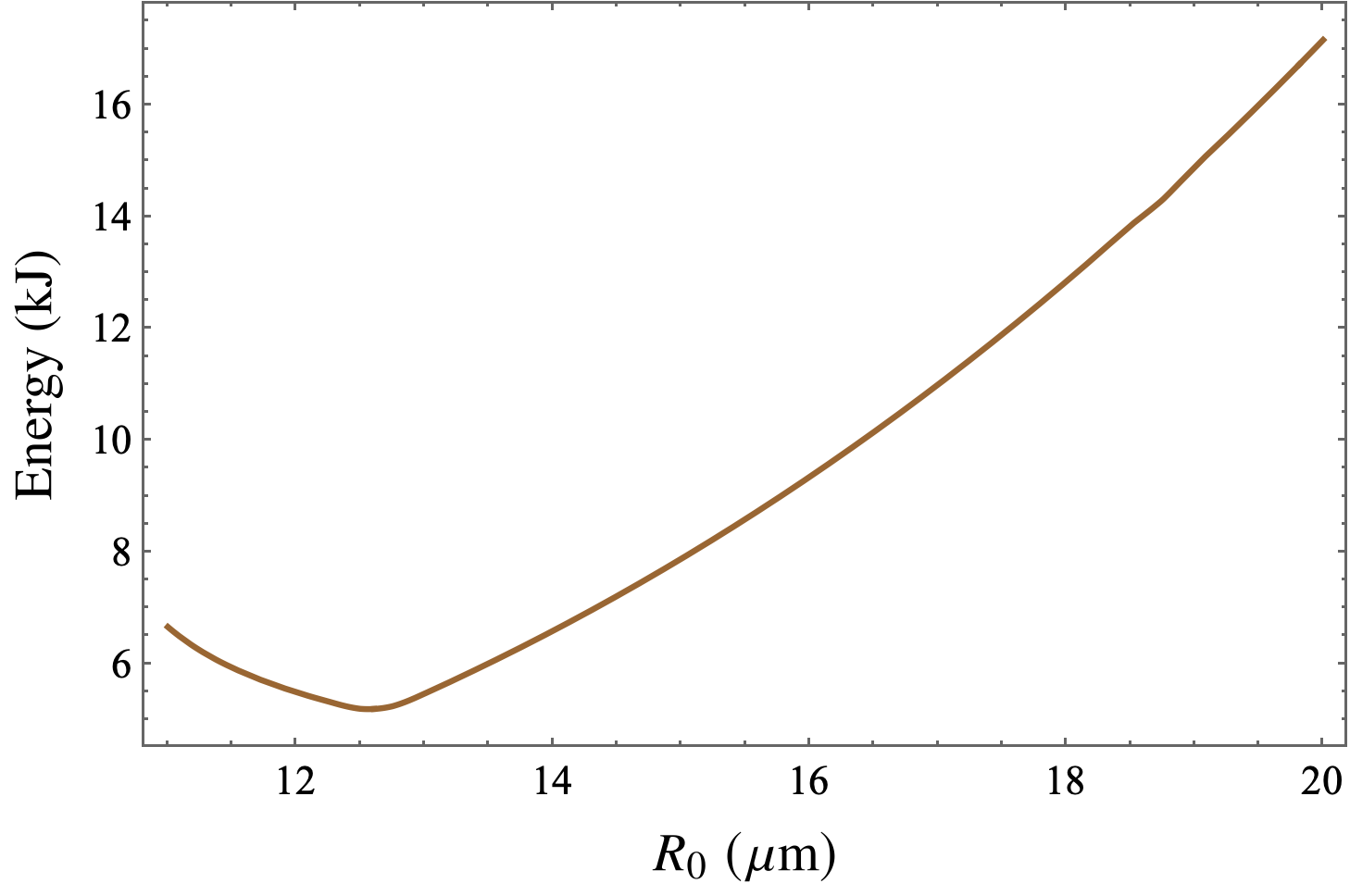}
\caption{Ignitor energy for 100 ps constant power pulse vs.\ initial hotspot radius for $\rho_0=100\text{ g/cm}^3$ and $\ln{\Lambda}=3.7$.}
\label{fig:100gcc}
\end{figure}

\begin{figure}[h]
\includegraphics[width=8cm]{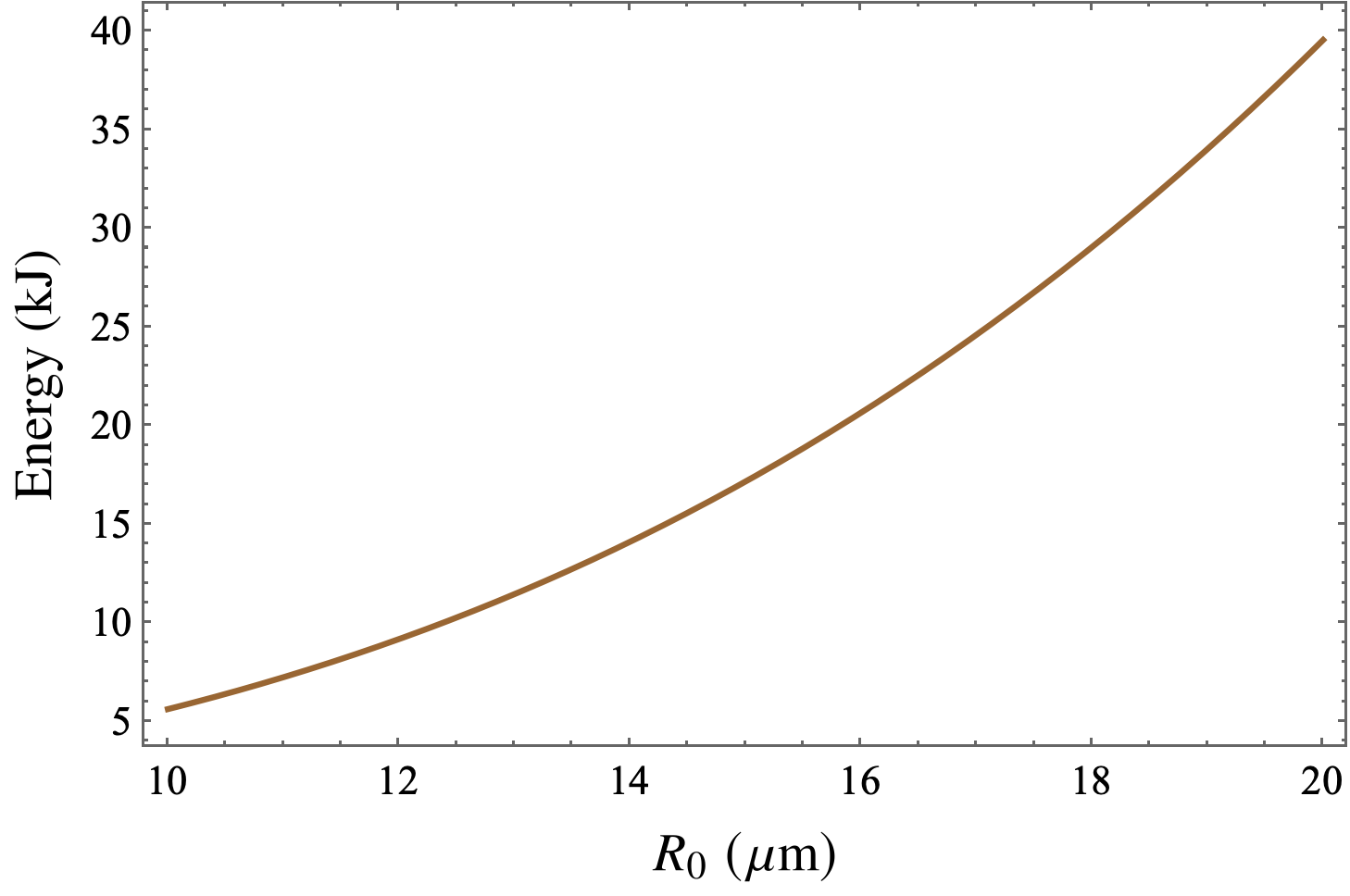}
\caption{Ignitor energy for 100 ps constant power pulse vs.\ initial hotspot radius for $\rho_0=200\text{ g/cm}^3$ and $\ln{\Lambda}=3.3$.}
\label{fig:200gcc}
\end{figure}

\begin{figure}[h]
\includegraphics[width=8cm]{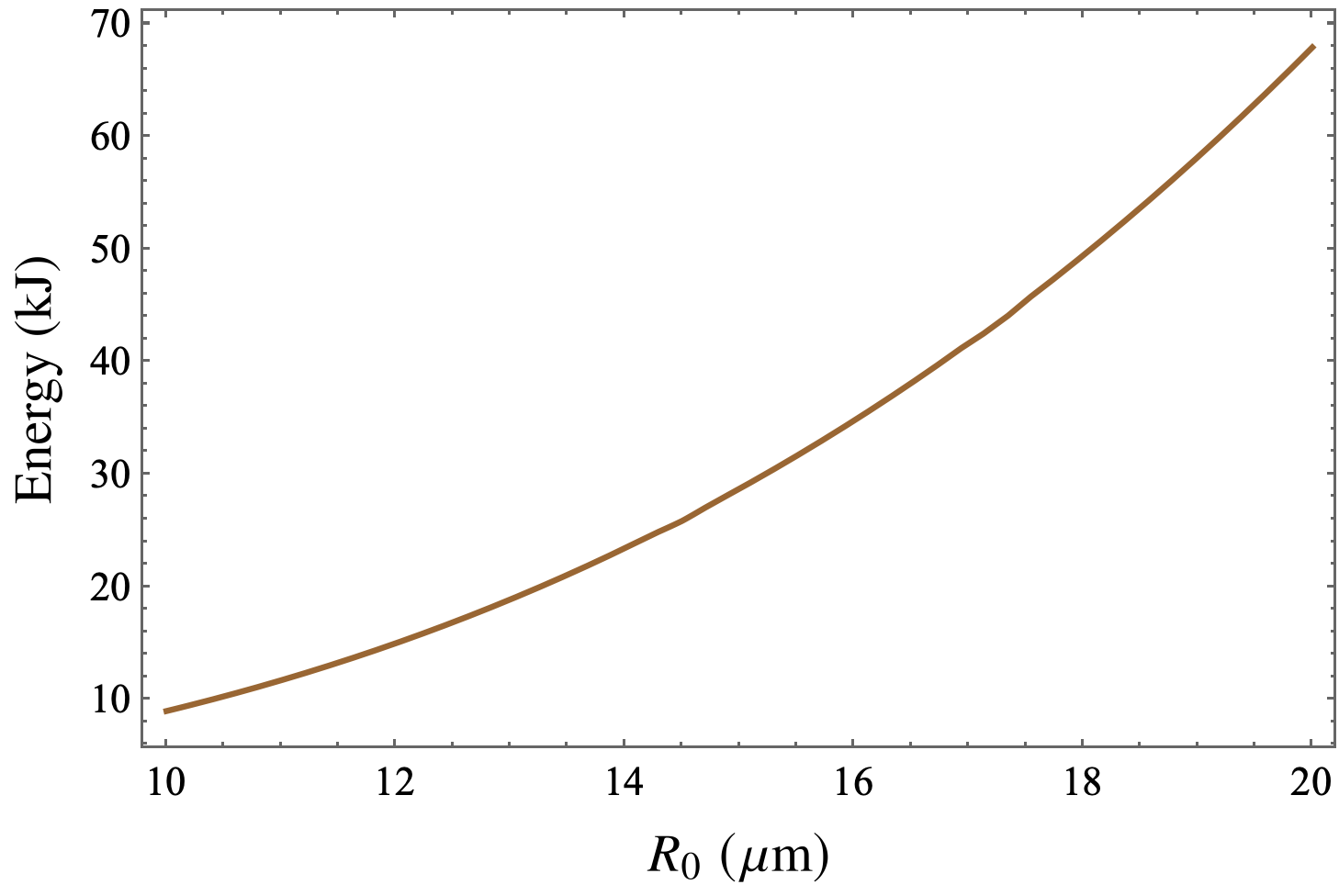}
\caption{Ignitor energy for 100 ps constant power pulse vs.\ initial hotspot radius for $\rho_0=300\text{ g/cm}^3$ and $\ln{\Lambda}=3$.}
\label{fig:300gcc}
\end{figure}

Now, for the bulk of our analysis, we investigate hotspot performance for constant power pulses given a magnetic field strength of $B_0=30\text{ kT}$ and a pulse length of $100\text{ ps}$. For ignition, we use a heuristic based on curve 1 in Fig.\ 2 in Slutz, Herrmann, and Vesey, et al. \cite{10.1063/1.3333505}. We are in the part of the curve where $R/R_\alpha$ ranges from 0.1 to 1.0 which is approximately linear, so we take as a condition for ignition $T_{ih}>10\text{ keV}$ and
\begin{equation}
    \log_{10}\left(\frac{R}{R_\alpha}\right)>-\frac{1}{1-\log_{10}(3)}\left(\log_{10}(\rho R)-\log_{10}(0.03)\right)
\end{equation}
where we use the numerical values of these quantities in cgs units. For a final hotspot expansion ratio $x=x_f$, this becomes approximately
\begin{equation}
    \log_{10}\left(\frac{R_0}{0.0009x_f}\right)>-1.91\left(\log_{10}\left(\frac{\rho_0 R_0}{x_f}\right)+1.52\right),
\end{equation}
so we prematurely halt integration if we hit an expansion ratio $x_c$ of
\begin{equation}
    x_c=111\rho_0^{0.66}R_0,
\end{equation}
because past $x=x_c$ ignition is no longer possible. For a hotspot of initial radius $R_{0}$, we set its length to be $L=5R_{0}$. We consider the densities $50\text{ g/cm}^3$, $100\text{ g/cm}^3$, $200\text{ g/cm}^3$, and $300\text{ g/cm}^3$ and plot the minimum ignitor energy required for ignition vs.\ initial hotspot radius in Figs. \ref{fig:50gcc}, \ref{fig:100gcc}, \ref{fig:200gcc}, and \ref{fig:300gcc} respectively. Some significant sets of parameters which achieved ignition are summarized in Table \ref{tab:table1}.

\begin{table}
\caption{\label{tab:table1}We summarize some significant sets of parameters for constant power pulses which achieved ignition here. Note that we used cylinder lengths $L=5R_0$.}
\begin{ruledtabular}
\begin{tabular}{ccccc}
 $\rho_0$ & $R_0$ & L & Pulse Length & Ignitor Energy \\
\hline
\\[-0.97em]
50 $\text{g/cm}^3$ & 17.75 \textmu m & 88.75 \textmu m & 100 ps & 6.41 kJ \\
100 $\text{g/cm}^3$ & 12.5 \textmu m & 62.5 \textmu m & 100 ps & 5.19 kJ \\
200 $\text{g/cm}^3$ & 10 \textmu m & 50 \textmu m & 100 ps & 5.63 kJ \\
300 $\text{g/cm}^3$ & 10 \textmu m & 50 \textmu m & 100 ps & 8.95 kJ \\
100 $\text{g/cm}^3$ & 11 \textmu m & 55 \textmu m & 60.24 ps & 4.01 kJ
\end{tabular}
\end{ruledtabular}
\end{table}

\section{\label{sec:discussion}Discussion}

Our best results are seen in the $\rho_0=\text{100 g/cm}^3$ case, with ignition achieved using a 5.19 kJ, 100 ps constant power pulse at an initial hotspot radius of 12.5 \textmu m. Lower energies are possible, but they come at the expense of higher power. For example, the $\rho_0=\text{100 g/cm}^3$ case could also ignite using a 4.01 kJ, 60.2 ps pulse at an initial hotspot radius of 11 \textmu m. The $\text{200 g/cm}^3$ case achieved ignition using a 5.63 kJ, 100 ps pulse at an initial hotspot radius of 10 \textmu m and could have done even better at lower initial hotspot radii, but smaller hotspots come with the tradeoff of greater difficulty focusing the ignitor beam into the hotspot. It may be that this is not as much of an issue in MagLIF due to the strong axial magnetic fields helping to collimate the ignitor beam electrons, but future research into this is required.

Applied magnetic fields of $\sim 5\text{ kT}$ have been considered for fast ignition to collimate ignitor electrons, dramatically reducing the required ignitor energy \cite{10.1063/1.4739294}. One difficulty of doing magnetized fast ignition in laser ICF is that charged particles from the ignitor beam are mirror reflected as they encounter increasing axial magnetic fields as they travel into the plasma, a problem which can be addressed through hydro design \cite{10.1063/1.4739294}. We propose another resolution to this problem which could be made possible by MagLIF: magnetizing the guiding cone where the ignitor electrons are born. While in laser ICF the cone must be able to survive being ablated by the X-ray burst from the hohlraum, in MagLIF there is no such X-ray burst which should relax design constraints on the cone, perhaps allowing for higher resistivity which would allow for diffusion of the flux compressed field into the cone. Even if we are not able to increase the resistivity of the cone, a MagLIF compression takes much longer than a laser ICF compression, which gives more time for the field to resistively diffuse into the cone anyways. Magnetizing the guiding cone not only solves the mirror reflection problem, but also provides another critical benefit in that it collimates the ignitor electrons right from birth. One of the fundamental obstacles to making fast ignition work is that there is an inevitable standoff distance between the cone and the hotspot caused by a high pressure jet flowing from the hotspot and impacting the cone; ignitor electrons diverge over this standoff distance, a disaster for energy deposition \cite{M_Tabak_2010}. If the ignitor electrons are collimated from birth, they will not diverge over the standoff distance, solving a fundamental problem with fast ignition. We additionally propose that there may be ways to have charged particles from the ignitor beam use mirror fields that they encounter to slow themselves down and increase their perpendicular energy, which would help energy deposition. Both these proposals are speculative and future research into them is required. Furthermore, MagLIF's large aspect ratio provides plenty of length for ignitor electrons to deposit their energy (see $L$ column in Table \ref{tab:table1}). This combination of ignitor electron collimation and large deposition length means that we could potentially achieve very good ignitor energy deposition in MagLIF.

There are a couple of serious objections that could be made to this model. One objection is the assumption of no alpha heating, despite alpha heating potentially significantly reducing the energy required for ignition. This is a conservative assumption, and including alpha heating would simply serve to save on ignitor energy. Another objection is the assumption of a large aspect ratio and with it the neglecting of axial losses. The assumption of a large aspect ratio is in a sense conservative because if we were able to somehow do something clever (such as the aforementioned proposal of exploiting magnetic mirroring) to deposit the ignitor energy within a short distance, then it would of course dramatically reduce the required ignitor energy; however, in general the strong axial magnetic fields and lower fuel density in MagLIF will naturally result in much higher hotspot aspect ratios than in laser ICF due to collimation of ignitor beam electrons and reduced stopping power respectively (although there needs to be future research into optimizing the ignitor electron energy spectrum to achieve desired hotspot geometries and energy deposition patterns). Furthermore, the results presented here should not be dismissed even for smaller aspect ratios. This is because as the hotspot expands radially, the magnetic field configuration of the hotspot forms a magnetic mirror with mirror ratio $x^2$. While the DT fuel ions are too collisional to benefit from kinetic mirror trapping, the mirror configuration is still beneficial in multiple ways. Firstly, there is a radial component of the magnetic field that insulates the hotspot from axial thermal conduction losses. Secondly, this is a small effect, but the mirror shaped magnetic fields provide axial confinement from magnetic pressure \cite{10.1063/5.0207561}. Thirdly, unlike for the fuel ions, there is in fact mirror trapping of the $3.5 \text{ MeV}$ fusion alphas as confirmed in simulations \cite{10.1063/1.4816813}. For a magnetic field of strength $30 \text{ kT}$, a fusion alpha has a gyroradius of 9 microns and is therefore magnetized. The distance of closest approach of a fusion alpha with DT fuel ions is $8.2\times 10^{-16} \text{ m}$, which even at a number density of $10^{25}\text{ cm}^{-3}$ corresponds to a mean free path of a whopping $4.7 \text{ cm}$. Therefore, fusion alphas are effectively collisionless (in the sense that they do not pitch-angle scatter) while slowing down on electrons in the hotspot, so their magnetic moment is conserved. Together, these three effects act to mitigate axial losses even for lower aspect ratio hotspots. The neglecting of axial thermal losses may seem suspicious despite the sides of the cylinder having much greater surface area than the ends: high magnetic fields suppress thermal losses out the sides making end losses the dominant thermal loss mechanism, and furthermore, the end surface area is significant, being on the order of the surface area of a typical laser ICF fast ignition hotspot. However, the neglecting of axial thermal losses is still justified for the following reasons: firstly, the large aspect ratio means that the temperature gradient for end losses is small (much less than the temperature gradients in laser ICF), secondly, as mentioned earlier there appears a radial component of the magnetic field which thermally insulates the hotspot axially, and finally, any axial thermal losses simply serve to elongate the hotspot axially, so axial thermal losses can be "accounted" for by assuming a long hotspot shape (which we do).

Besides adding alpha heating, better results could also be obtained by considering more sophisticated pulse shapes other than constant power pulses and by considering intermediate densities in between the ones considered here. Another effect that could be considered for future research is shear flow enhanced reactivity \cite{10.1063/5.0285620,5nll-y8rx,10.1063/5.0295335}.

Overall, the results presented here demonstrate a fast ignition regime that lies outside the parameter space traditionally considered feasible for fast ignition in laser ICF, with ignition achieved using a long-pulse ($>100\text{ ps}$), low-energy ($<10\text{ kJ}$) ignitor. This is a relaxation on ignitor power requirements that now puts fast ignition within plausible reach of existing plans for fast ignition reactors, including FIREX-II with a 50 kJ, 10 ps ignitor \cite{Motojima_2007} and HiPER with a 100 kJ, 15 ps ignitor \cite{article}. Note that if we were required to deliver 20 kJ to the fuel in 20 ps, which as mentioned earlier is what would be required for fast ignition in laser ICF at the standard density of $300\text{ g/cm}^3$, then for a coupling of 10\% we would need a 200 kJ, 10 PW ignitor, and both FIREX-II and HiPER would be wholly inadequate for that purpose. On the other hand, if we were only required to deliver $\gtrsim 5\text{ kJ}$ to the fuel in 100 ps as the results in this paper indicate, then both FIREX-II and HiPER would be massively overkill.

Another reason why fast ignition may be fundamentally better suited for MagLIF than laser ICF is repetition rate. HiPER was envisioned to have a repetition rate of 10 Hz \cite{article}, a highly challenging task. MagLIF naturally lends itself to higher yields than laser ICF due to magnetic fields allowing for ignition at lower pressure \cite{alexander2025affordablemanageablepracticalscalable}, better coupling of driver energy to the fuel \cite{alexander2025affordablemanageablepracticalscalable}, and more efficient heating of the fuel per ion (lower densities result in lower bremsstrahlung losses per ion and magnetic fields suppress thermal losses). Higher yields, especially if we could reach the multi-gigajoule yields \cite{PhysRevLett.108.025003,10.1063/1.4941100} that MagLIF allows for in principle, could substantially reduce repetition rate requirements and any corresponding strain on the fast ignition laser system.

Even if we compress cold from cryogenic fuel, the minimum achievable stagnation pressure is limited by the Fermi degeneracy pressure. A comparison between the Fermi degeneracy pressure at various densities and a reference pressure representing a high yield compression heating setup achieving a stagnation density and temperature of $5\text{ g/cm}^3$ and $10\text{ keV}$ is shown in Table \ref{tab:table2}. We see that even though MagLIF can already ignite at much lower pressures than laser ICF, fast ignition still offers the potential for substantial reductions in stagnation pressure. In particular, the Fermi pressure for $\rho_0=50\text{ g/cm}^3$ is 1.47 Gbar while the Z machine has reached experimental stagnation pressures of $\sim 1.8\text{ Gbar}$ \cite{10.1063/5.0087115} to $\gtrsim 2\text{ Gbar}$~\cite{10.1063/5.0244304}, demonstrating the possibility of high yield MagLIF even at stagnation pressures on the order of those achievable by the Z machine. One issue is magnetic pressure, with a 30 kT magnetic field having a magnetic pressure of 3.58 Gbar, so future research on optimally trading off magnetic pressure with degeneracy pressure is required. Higher densities require much higher stagnation pressures due to Fermi degeneracy pressure scaling with $\rho_0^{5/3}$ but presumably offer higher yields, so there should also be future research on optimally trading off stagnation pressure with yield.

\begin{table}
\caption{\label{tab:table2}Comparison of Fermi degeneracy pressure $P_d$ at various densities to that of a reference pressure $P_0$ at $5\text{ g/cm}^3$ and $10\text{ keV}$}
\begin{ruledtabular}
\begin{tabular}{cc}
 $\rho_0$ & $P_d/P_0$ (\%)
\\
\hline
\\[-0.97em]
50 $\text{g/cm}^3$ & 3.82\% \\
100 $\text{g/cm}^3$ & 12.16\% \\
200 $\text{g/cm}^3$ & 38.60\% \\
300 $\text{g/cm}^3$ & 75.87\%
\end{tabular}
\end{ruledtabular}
\end{table}

Due to fast ignition requiring compression at cold temperatures, the plasma will not be conductive enough for the frozen-in law to hold, and the conductivity of the liner will have to be relied on for flux compression. This is feasible because the liner is highly conductive, with the seed field coils on the Z machine being required to have a rise time on the order of $\sim 1\text{ ms}$ \cite{10.1063/1.3333505} (i.e. much longer than the compression time) to give the field enough time to diffuse into the liner. The conductivity of the liner varies during the compression, with the liner heating up (initially decreasing conductivity but then later increasing it) and thickening (increasing conductivity), but overall the liner should be able to remain conductive enough to do flux compression. Note that our assumption that the frozen-in law holds during the ignitor pulse becomes approximately true as the fuel heats up and is also a conservative assumption because any resistive diffusion of magnetic field into the hotspot would be helpful for ignition. Unavoidable sources of heating during compression include shock heating and ohmic heating from current flowing through the liner. While shock heating should be minimized as much as possible to keep stagnation pressure low, ohmic heating could possibly be beneficial because it acts on the outer regions of the fuel so it may be able to raise the central density at stagnation through pressure balance. This ohmic heating can be increased through the addition of resistive heating elements, so there should be future research on if we can actually achieve higher central densities by heating the outer regions of the fuel. Magnetic fields are excellent at facilitating burn propagation parallel to the field lines \cite{10.1063/5.0197817}, so we expect that any burn started from injecting fast electrons at one end of the MagLIF cylinder will propagate axially throughout the entire cylinder for high yield, although future research into optimizing burn propagation and cylinder length is required.

This research was supported by DOE contract grant DE-AC02-09CH11466 (PPPL) and also by the Center for Magnetic Acceleration, Compression, and Heating (MACH), part of the U.S. DOE-NNSA Stewardship Science Academic Alliances Program under Cooperative Agreement No. DE-NA0004148. The United States Government retains a non-exclusive, paid-up, irrevocable, world-wide license to publish or reproduce the published form of this manuscript, or allow others to do so, for United States Government purposes.

\nocite{*}
\bibliography{aipsamp}

\end{document}